\documentclass[universe,article,accept,moreauthors,pdftex]{Definitions/mdpi} 

\usepackage{setspace}
\usepackage{graphicx}

\usepackage{amsmath,amssymb,amsfonts,amsthm,mathrsfs}
\usepackage{enumerate} 
\firstpage{1} 
\pubvolume{00}
\issuenum{1}
\articlenumber{5}
\pubyear{2022}
\copyrightyear{2022}
\history{Received: date; Accepted: date; Published: date}

\Title{A covariant polymerized scalar field in semi-classical loop quantum gravity}

\Author{Rodolfo Gambini $^{1}$, Florencia Benítez$^{2}$ and Jorge Pullin $^{3,}$*}

\AuthorNames{Firstname Lastname, Firstname Lastname and Firstname Lastname}

\address{%
$^{1}$ \quad Instituto de Física, Facultad de Ciencias, Universidad de la República, Montevideo, 11400, Uruguay; rgambini@fisica.edu.uy\\
$^{2}$ \quad Instituto de Física, Facultad de Ingeniería, Universidad de la República, Montevideo, 11400, Uruguay; fbenitez@fisica.edu.uy\\
$^{2}$ \quad Department of Physics and Astronomy, Louisiana State University,
  Baton Rouge, LA 70803-4001, USA}

\corres{Correspondence: pullin@lsu.edu}

\abstract{
We propose a new polymerization scheme for scalar fields coupled to gravity. It has the advantage of being a (non-bijective) canonical transformation of the fields and therefore ensures the covariance of the theory. We study it in detail in spherically symmetric situations and compare to other approaches.}

\begin{document}
\section{Introduction}

``Polymerization'' is a common procedure for constructing candidates for semi-classical theories stemming from loop quantum gravity. The idea is that in the Hilbert space \cite{asle} commonly used in loop representations of diffeomorphism invariant theories like general relativity, the connection is not a well defined variable. However, its holonomy is. When matter fields are present, or when one is working in situations with reduced symmetry, some of the variables that are connections in the full theory may become scalars, requiring point holonomies. Yet, to try to mimic the behaviors one would see in the full theory, or to make the inclusion of matter compatible with the Hilbert spaces of interest, one usually considers inner products in which certain variables are not well defined although their exponentials are. To construct the equations of the theory one therefore replaces the variables in question by their exponentials, or, since one is interested in real variables, the substitution tends to be of the form $x\to \sin(k x)/k$ where $x$ is the variable in question and $k$ is known as the polymerization parameter. It is clear that in the limit $k\to 0$ one recovers the original classical theory from the ``polymerized'' one that attempts to capture quantum corrections.  The polymerization parameter $k$ plays the role of the length of the loop along which one would compute a holonomy if the variable in question had been a connection. Sometimes the exponentiated quantity, in the case where the variable is not a connection, is known as ``point holonomy''. This type of construction has been used widely in loop quantum cosmology \cite{lqc} and also in spherically symmetric situations. It is also understood as a non-standard representation of the canonical commutation relations and can be applied in ordinary systems in quantum mechanics \cite{covuza}. The use of point holonomies has also been proposed in the full theory to represent scalar fields \cite{scalar}. 

The use of polymerizations has been criticized for the lack of covariance. Since one is typically working in a canonical context the polymerization is applied to spatial variables. The construction is generically slicing-dependent \cite{bojo5}.

In this paper we would like to propose a new polymerization procedure for scalar variables and apply it to the case of a scalar field coupled to spherically symmetric gravity. The novelty is that the polymerization procedure is a non-bijective canonical transformation on the variables of classical general relativity coupled to a scalar field. As a result, its application appears more compatible with covariance. At first this may appear surprising. If it is a canonical transformation one is essentially dealing with the same theory. How could this be capturing non-trivial quantum corrections to the theory? The answer is that the canonical transformation is non-invertible and therefore the resulting theory is not unitarily equivalent to the non-holonomized one \cite{Moshinsky,Anderson,Deenen}. The canonical transformation leads to a theory that can be viewed as the semiclassical theory stemming from a non-standard representation of the Weyl algebra. As we shall see, the construction for the scalar field is general, but for the geometrical variables we do not know a general polymerization procedure yet. We will restrict to study in detail the polymerization  in the context of effective spherically symmetric gravity to make it concrete.

Recently, the effective theory resulting from a standard polymerization of a gauge fixed theory of gravity coupled to a scalar field in spherical symmetry was studied \cite{choptuikus} to address the semiclassical corrections of gravitational collapse. This system had been studied in classical general relativity by Choptuik \cite{choptuik}. We will show that the results of that analysis do not change significantly with the polymerization proposed in this paper.

\section{Spherically symmetric general relativity in new variables}

We recall the basics of spherically symmetric gravity minimally coupled to a massless scalar field \cite{Gambini:2009ie}. The classical variables are the triads in the radial and transverse directions and their canonically conjugate momenta $E^x,E^\varphi, K_x,K_\varphi$. Their relations to the usual metric variables are,
\begin{equation}
ds^{2}=\varLambda^{2}dx^{2}+R^2d\Omega^{2}
\end{equation}
with,
$\Lambda=E^\varphi/\sqrt{\vert
  E^x\vert}$, $R^2=\vert E^x\vert$ and to the extrinsic
curvatures $K_{xx}=-{\rm sign}(E^x) \left(E^\varphi\right)^2
K_x/\sqrt{\vert E^x\vert}$ and $K_{\theta \theta}= -\sqrt{\vert
  E^x\vert} K_\varphi/(2\beta)$ with $\beta$ the Immirzi parameter. Here $ds^2$ is the spatial metric, the space-time metric is reconstructed in the usual way introducing a lapse $N$ and shift $N^x$.

The total Hamiltonian is,

\begin{eqnarray}
H_{T}&=&\int dx\left\{ N^{x}\left(\left(E^{x}\right)'K_{x}-E^{\varphi}\left(K_{\varphi}\right)'-8\pi P_{\phi}\phi'\right)\right.\nonumber\\
&&\left.+N\left(-\frac{E^{\varphi}}{2\sqrt{\vert E^x\vert}}-2\sqrt{\vert  E^x\vert}K_{\varphi}K_{x}-\frac{K_{\varphi}^{2}E^{\varphi}}{2\sqrt{\vert  E^x\vert}}+\frac{\left(\left(E^{x}\right)'\right)^{2}}{8\sqrt{\vert  E^x\vert}E^{\varphi}}\right.\right.\nonumber\\
&&\left.\left.
-\frac{\sqrt{\vert E^x\vert}\left(E^{x}\right)'\left(E^{\varphi}\right)'}{2\left(E^{\varphi}\right)^{2}}+\frac{\sqrt{\vert  E^x\vert}\left(E^{x}\right)''}{2E^{\varphi}}+\frac{2\pi P_{\phi}^{2}}{\sqrt{\vert  E^x\vert}E^{\varphi}}+\frac{2\pi \left(\vert  E^x\vert \right)^{3/2}\left(\phi'\right)^{2}}{E^{\varphi}}\right)\right\} .
\end{eqnarray}
where prime denotes 
${\partial }/{\partial x}$.

\bigskip{}
The Hamiltonian and diffeomorphism constraint satisfy the usual algebra with structure functions. This makes quantization difficult. However, 
redefining the shift and lapse in the following way,
\begin{equation}
\bar{N}^{x}=N^{x}+\frac{2NK_{\varphi}\sqrt{\vert E^x\vert}}{\left(E^{x}\right)'}
\end{equation}

\begin{equation}
\bar{N}=\frac{E^{\varphi}N}{\left(E^{x}\right)'}
\end{equation}

\bigskip{}

one has that,\bigskip{}

\begin{equation}
H_{T}=\int dx\left(\bar{N}^{x} D_x+\bar{N} H \right)
\end{equation}


\bigskip{}

\bigskip{}

With the diffeomorphism constraint,
\begin{equation}
D_x=\left(E^{x}\right)'K_{x}-E^{\varphi}\left(K_{\varphi}\right)'-8\pi P_{\phi}\phi'
\end{equation}
and Hamiltonian constraint

\begin{equation}
H=\left[\sqrt{\vert E^x\vert}\left(\frac{\left(\left(E^{x}\right)'\right)^{2}}{4\left(E^{\varphi}\right)^{2}}-1-K_{\varphi}^{2}\right)\right]^{'}-\frac{2K_{\varphi}\sqrt{\vert E^x\vert}\phi'P_{\phi}}{E^{\varphi}}+\frac{2\pi\left(E^{x}\right)'P_{\phi}^{2}}{\sqrt{\vert E^x\vert}\left(E^{\varphi}\right)^{2}}+\frac{2\pi \left (\vert  E^x\vert\right)^{3/2} \left(E^{x}\right)'\left(\phi'\right)^{2}}{\left(E^{\varphi}\right)^{2}} \label{eq:14-1}
\end{equation}
such that the latter has an Abelian algebra with itself. In the vacuum case this paved the way for the complete quantization of the model\cite{gapu}. With the scalar field present, the term involving the derivative interferes with  the Abelianization involved in promoting the algebra to a consistent quantum one.  This led us in a previous paper \cite{choptuikus} to consider the polymerized version of a totally gauge fixed form of the theory following closely what was done by Choptuik. The choice $E^x=x^2$, $K_{\varphi}=0$ leads to a reduced theory where only the scalar field needs to be polymerized. A question remains when one follows this quantization procedure. Will different gauge and polymerizations choices lead to an equivalent quantum theory?

\section{A new covariant polymerization}

We will apply the following canonical transformations for the scalar field, $\phi$, the curvature, $K_{\varphi}$, and their canonical momenta, $P_{\phi}$ and $E^{\varphi}$,

\begin{equation}
\begin{array}{cc}
\phi\mapsto\frac{\sin\left(k\varphi\right)}{k}, & P_{\phi}\mapsto\frac{P_{\varphi}}{\cos\left(k\varphi\right)}\\
K_{\varphi}\mapsto\frac{\sin\left(\rho K_{\varphi}\right)}{\rho}, & \;E^{\varphi}\mapsto\frac{E^{\varphi}}{\cos\left(\rho K_{\varphi}\right)}
\end{array}\label{eq:8}
\end{equation}
where $k$ and $\rho$ are the polymerization parameters for the field and curvature, respectively. Canonical transformations in the spherically symmetric context have also been considered in \cite{tibrewala}. It should be noted that the canonical transformation is not bijective in all of phase space. It therefore is a proper canonical transformation provided $\cos{(\rho K_\varphi)} \ne 0$.  The emergence of new physics is possible if one considers ranges where the transformation is not invertible.

Applying the canonical transformation we have that,
\bigskip{}

\begin{eqnarray}
H&=&\frac{\sqrt{E^{x}}\left(E^{x}\right)'\left(E^{x}\right)''\cos^{2}\left(\rho K_{\varphi}\right)}{2\left(E^{\varphi}\right)^{2}}+\frac{\left(\left(E^{x}\right)'\right)^{3}\cos^{2}\left(\rho K_{\varphi}\right)}{8\left(E^{\varphi}\right)^{2}\sqrt{E^{x}}}\nonumber\\
&&-\frac{\sqrt{E^{x}}\left(\left(E^{x}\right)'\right)^{2}\left(E^{\varphi}\right)'\cos^{2}\left(\rho K_{\varphi}\right)}{2\left(E^{\varphi}\right)^{3}}
-\frac{\sqrt{E^{x}}\left(\left(E^{x}\right)'\right)^{2}\rho\left(K_{\varphi}\right)'\sin\left(\rho K_{\varphi}\right)\cos\left(\rho K_{\varphi}\right)}{2\left(E^{\varphi}\right)^{2}}\nonumber\\
&&-\frac{\left(E^{x}\right)'}{2\sqrt{E^{x}}}\left(1+\left(\frac{\sin^{2}\left(\rho K_{\varphi}\right)}{\rho^{2}}\right)\right)-\frac{2\sqrt{E^{x}}\sin\left(\rho K_{\varphi}\right)\cos\left(\rho K_{\varphi}\right)\left(K_{\varphi}\right)'}{\rho}\nonumber\\
&&
-\frac{2\sin\left(\rho K_{\varphi}\right)\cos\left(\rho K_{\varphi}\right)\sqrt{E^{x}}P_{\varphi}\varphi'}{\rho E^{\varphi}}+\frac{2\pi\left(E^{x}\right)'\cos^{2}\left(\rho K_{\varphi}\right)P_{\varphi}^{2}}{\sqrt{E^{x}}\left(E^{\varphi}\right)^{2}\cos^{2}\left(k\varphi\right)}\nonumber\\
&&+\frac{2\pi\sqrt{E^{x}}E^{x}\left(E^{x}\right)'\left(\varphi'\right)^{2}\cos^{2}\left(k\varphi\right)\cos^{2}\left(\rho K_{\varphi}\right)}{\left(E^{\varphi}\right)^{2}}.\label{eq:13-1}
\end{eqnarray}

\bigskip{}

The diffeomorphism constraint is unchanged, therefore so it is its algebra with itself and the Hamiltonian. Moreover, one can check that the Hamiltonian remains Abelian with itself,

\begin{equation}
\left\{ H\left(\bar{N}(x)\right);H\left(\bar{M}(y)\right) \right\}=0.
\end{equation}

So the resulting theory has the same constraint algebra as in referece \cite{gapu}. 

\section{Relation to other polymerized approaches}

As we mentioned, gravity with spherical symmetry coupled to a scalar field was recently studied in the context of a semi-classical loop quantum gravity analysis of the phenomena discussed by Choptuik \cite{choptuikus}. In that work, we used the same gauge fixing as Choptuik \cite{choptuik} had considered in his original analysis in classical general relativity. In it one takes the usual Schwarzschild coordinates for the exterior of a black hole, plus an additional condition on the extrinsic curvature. In terms of our variables, this corresponds to $E^{x}=x^{2}$, $K_{\varphi}=0$. Let us see how the equations with the new polymerization we present in this paper look like in that gauge choice:

\begin{equation}
\frac{N'}{N}-\frac{\left(E^{\varphi}\right)'}{E^{\varphi}}+\frac{2}{x}-\frac{\left(E^{\varphi}\right)^{2}}{x^{3}}=0\label{eq:1-2}
\end{equation}

\begin{equation}
\frac{\left(E^{\varphi}\right)'}{E^{\varphi}}-\frac{3}{2x}+\frac{\left(E^{\varphi}\right)^{2}}{2x^{3}}-2\pi x\left(\frac{\left(P_{\varphi}\right)^{2}}{x^{4}\cos^{2}\left(k\varphi\right)}+\left(\varphi'\right)^{2}\cos^{2}\left(k\varphi\right)\right)=0\label{eq:2-3}
\end{equation}

\begin{equation}
\dot{\varphi}=\frac{4\pi NP_{\varphi}}{E^{\varphi}x\cos^{2}\left(k\varphi\right)}
\end{equation}

\begin{eqnarray}
\dot{P_{\varphi}}&=&-\frac{4\pi NP_{\varphi}^{2}}{E^{\varphi}x}\frac{k\sin\left[k\varphi\right]}{\cos^{3}\left(k\varphi\right)}+\frac{4\pi x^{2}}{E^{\varphi}}
\left[\left(\frac{3NE^{\varphi}-xN\left(E^{\varphi}\right)'+N'E^{\varphi}x}{E^{\varphi}}\right)\varphi'\cos^{2}\left(k\varphi\right)\right.\nonumber\\
&&\left.+xN\varphi''\cos^{2}\left(k\varphi\right)-xNk\left(\varphi'\right)^{2}\cos\left(k\varphi\right)\sin\left(k\varphi\right)\right]
\end{eqnarray}

\begin{equation}
K_{x}=\frac{-4\pi P_{\varphi}\varphi'}{x}\label{eq:21}
\end{equation}

\bigskip{}

Comparing with the more traditional polymerization we considered in \cite{choptuikus}, the last three equations are modified. The first term in the last equation is new and the cosines in the denominator of the second and third equations were absent. These terms can potentially make a significant difference in some regions of phase space, like close to singularities as noted in reference \cite{choptuikus}.

To try to test this, we conducted numerical simulations like the ones in our previous paper \cite{choptuikus} to determine the scaling law of the mass of the final black hole formed by the collapse of a scalar field as a function of a parameter in the initial data. The figure shows the comparison for the rather unnaturally large value $k=1$ of the polymerization parameter. As can be seen, it coincides with general relativity ($k=0$). Of course, since the Choptuik phenomena is determined by the exterior geometry of the black hole where the fields are weak, it is perhaps not surprising that the polymerization yields the same result.  On the other hand, it provides evidence that the results found in our previous paper do not have significant slicing dependence, as they differ little from the new polymerization which does not depend on slicings. It should be noted that, for small black holes, the Choptuik phenomena do involve regions of large curvature immediately outside the horizon and results may therefore change upon polymerization. 

\begin{figure}
    \centering
    \includegraphics[width=11cm]{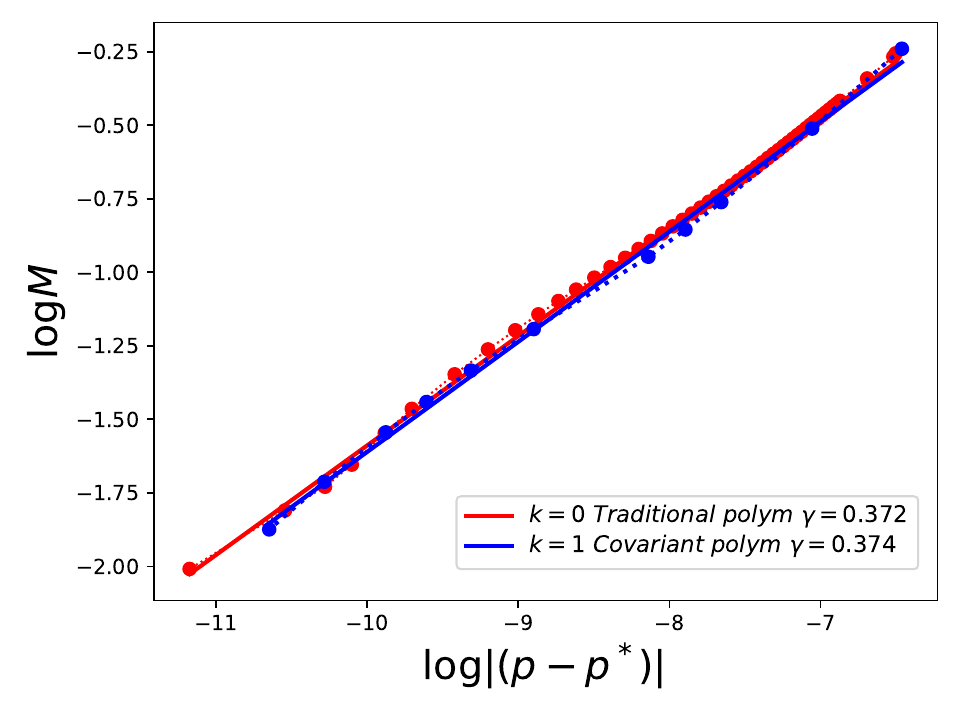}
    \caption{The scaling of the black hole mass observed first by Choptuik \cite{choptuik} for the collapse of spherically symmetric massless scalar field. Here we depict it for general relativity ($k=0$). The other curve is  for the polymerization parameter value $k=1$, which in practice is a very exaggerated value, in order to make any possible discrepancies with classical general relativity more visible. Here, $\gamma$ is the universal exponent for the mass law of the black hole (see \cite{choptuikus}). As can be seen both theories agree. The dots are numerical results and the solid lines are least square fits.}
    \label{fig:my_label}
\end{figure}

\section{Discussion}

We have introduced a new polymerization for scalar fields coupled to gravity. This can be viewed as a non-standard representation of the Weyl algebra different than the one usually considered in loop quantum gravity coupled to scalar fields. It has the advantage that it is a canonical transformation from the original variables. That means that it preserves the constraint algebra and the covariance of the theory, which previous choices did not. As it is not-invertible in the whole of phase space it still allows to have the usual novel phenomena that loop quantizations introduce in regions where one expects general relativity not to be valid, like close to singularities. In particular it will admit a representation on a Hilbert space defined in terms of the Ashtekar Lewandowski measure \cite{asle}. Although we have only explored the implications of the covariant polymerization in the context of spherically symmetric scalar fields, it is possible that an analogue could be found for the full theory. This will require further investigations.

The reader might be curious about the question of singularity resolution
with the proposed polymerization. We have not made an exhaustive analysis
yet, but it appears that the same ingredients that have led to the resolution
of the singularity in spherical loop quantum gravity in previous polymerizations
(both the $\mu_0$ \cite{gapu} and $\bar{\mu}$ \cite{gapu2} quantizations) are present
here. Namely, that the parameterized Dirac observable corresponding to the
metric becomes complex for radii smaller than a given radius. This requires
removing points from the spin network in the region where the classical
singularity used to be to ensure self-adjointness of the metric. Those ingredients are still present in the current
polymerization. In phase space, one essentially has a ``bounce''
at the hypersurface at which $K_\varphi$ is maximum and the singularity cannot be reached. It should be noted that in previous treatments, the constraints implied both variables were bounded, so the additional changes implied here do not change things much in the quantum theory.

As was pointed in \cite{bojo}, the more balanced polymerization
introduced in this paper has the inconvenience of potentially introducing
significant quantum corrections in regions of small curvature where
the extrinsic curvature takes its maximum allowed value. However, in a full quantum theory of gravity,  reference frames
that require this type of curvatures will not be physically implementable since they require extrinsic curvatures exceeding the Planck scale. Covariance should be checked within implementable reference frames. In generally covariant theories reference frames have to be defined in terms of physical observables and all physical statements must be relational. This clearly limits which reference frames can be implemented. 
The problem might also be
addressed via an improved quantization, as was the potential
appearance of large corrections in regions of low curvature in the
original (``$\mu_0$'') version of loop quantum cosmology \cite{improved} and one could cover the complete space-time with implementable reference frames.   A related
proposal \cite{brizuela} considering  a family of effective modified constraints that satisfy Dirac's deformation algebra claims to be exempt from these issues. 

The Abelian Hamiltonian constraint given by our equation (\ref{eq:13-1}) together with the diffeomorphism constraint imply at
the effective level the original set of constraints. They can be derived from the Abelian set by substituting the rescaled Lagrange multipliers by their expression in terms of the original ones. By considering the algebra of the original constraints starting from the polymerized Poisson Brackets that we proposed, it is easy to prove that they satisfy the usual constraint algebra. As it has been proved in \cite{brizuela2} the general relativity constraint algebra with this polymerization implies the tensorial behavior of the metric components given by equations (13-16) of that paper. Thus we recover from our formulation the standard behavior of the metric components.

For this work we considered a polymerization with a constant parameter. In the future we may consider the case in which the polymerization parameter depends on the dynamical variables. This seem eminently feasible applying the results of \cite{improvedold} of the vacuum case.

\section*{Acknowledgements}
We wish to thank Martin Bojowald, Luis Lehner and Steve Liebling for discussions.
This work was supported in part by Grant NSF-PHY-1903799, NSF-PHY-2206557, funds of the Hearne Institute for Theoretical Physics, CCT-LSU,  Pedeciba, Fondo Clemente Estable FCE\_1\_2019\_1\_155865.

\end{document}